\pdfoutput=1
\documentclass[11pt]{article}
\usepackage{graphicx} 
\usepackage{amsmath}
\usepackage[margin=1in]{geometry}
\newcommand{\keywords}[1]{\textbf{Keywords:} #1}
\usepackage[backend=biber, style=apa]{biblatex}
\usepackage{amssymb}
\usepackage{xcolor}
\usepackage{subcaption}
\usepackage{caption}
\usepackage{booktabs}
\usepackage{csquotes}
\usepackage{bm}

\usepackage[most]{tcolorbox}
\usepackage{hyperref}

\newcounter{mybox} 
\renewcommand{\themybox}{\arabic{mybox}} 

\title{How brains build higher order representations of uncertainty}
\date{  }
\author{Megan A. K. Peters$^{*1,2,3,4,5}$, Hojjat Azimi Asrari$^{*1}$}
\date{\small{$^1$Department of Cognitive Sciences, University of California Irvine, Irvine, CA 92617 USA\\%
      $^2$Department of Logic \& Philosophy of Science, University of California Irvine, Irvine, CA 92617 USA\\
      $^3$Center for the Neurobiology of Learning \& Memory, University of California Irvine, Irvine, CA 92617 USA\\
      $^4$Center for Theoretical Behavioral Sciences, University of California Irvine, Irvine, CA 92617 USA\\
      $^5$Program in Brain, Mind, \& Consciousness, Canadian Institute for Advanced Research, Toronto, Ontario, M5G 1M1 Canada}}

\begin{document}

\maketitle

\noindent \small{$^*$ Corresponding authors: M.A.K.P. (megan.peters@uci.edu) and H.A.A. (hazimias@uci.edu)}

\newpage
\begin{abstract}
Higher-order representations (HORs) are neural or computational states that are ``about" first-order representations (FORs), encoding information not about the external world per se but about the agent's own representational processes—such as the reliability, source, or structure of a FOR. These HORs appear critical to metacognition, learning, and even consciousness by some accounts, yet their dimensionality, construction, and neural substrates remain poorly understood. Here, we propose that metacognitive estimates of uncertainty or noise reflect a read-out of ``posterior-like" HORs from a Bayesian perspective. We then discuss how these posterior-like HORs reflect a combination of ``likelihood-like" estimates of current FOR uncertainty and ``prior-like" learned distributions over expected FOR uncertainty, and how various emerging engineering and theory-based analytical approaches may be employed to examine the estimation processes and neural correlates associated with these highly under-explored components of our experienced uncertainty.

\end{abstract}

\newpage

\section*{Introduction}

As comprehensively discussed by Baker and colleagues \parencite{baker2022three}, the definition of a `neural representation' is hotly debated: ask three researchers (or three \textit{fields} of research), and you'll get three different answers \parencite{favela2023investigating, favela2025contextualizing,Vilarroya2017Neural,Machery2025Neural}. Here, we take at face value that one possible definition of neural representations is that they are more than just statistical covariation between patterns of neural response and relevant aspects of an observer's environment or mental processing \parencite{baker2022three,Ritchie2019Decoding}, instead reflecting the sorts of \textit{mental structures} that observers use to perceive, reason about, and engage with their environments \parencite{tarr2002visual}. This leads us to want to explore the \textit{kinds} of representations that might be relevant for an agent's behavior or cognitive capacities, and to then examine how neural correlates of such representations might be scientifically studied. 

In this paper, we first examine the difference between the more-commonly studied ``first-order representations" (FORs), which encode or represent features of the external world, and the less-commonly studied ``higher-order representations" (HORs), which encode properties of FORs such as their noise or evidentiary strength. While HORs are foundational to learning, introspection, and potentially consciousness, their computational structure and neural instantiation remain poorly understood. We then introduce and discuss the proposal that HORs about noise or uncertainty might represent these aspects of their target FORs specifically in multiple variants, following a general Bayesian-like framework: (1) by encoding the current unreliability or variability of a first-order signal, which we term \textit{likelihood-like HORs}; (2) by maintaining structured expectations the typical noise associated with certain stimuli or states, which we term \textit{prior-like HORs}; and (3) by combining these estimates into \textit{posterior-like} HORs about uncertainty. We synthesize recent theoretical and analytical solutions to accessing and characterizing each of these HOR variants, and discuss how they may be integrated to ultimately produce these posterior-like higher-order estimates of FOR uncertainty which may drive metacognitive assessments, learning, or conscious awareness.

\subsection*{First-order versus higher-order (neural) representations}

In much of the literature on neural representations, the target of study is representations that are ``about" aspects of the agent's environment: objects or features of the environment itself \parencite{baker2022three, tarr2002visual}, decision variables about that environment leading to behavioral outputs \parencite{gold2007neural}, memories \parencite{squire1991medial}, goals \parencite{miller2001integrative}, or actions the agent might take to achieve such goals \parencite{rizzolatti2004mirror,Thornton2024Neural}, for example. Here, we use ``about" in quotes to emphasize that the target of a mental representation -- that is, what it refers to (see also the Representational Theory of Mind; \cite{von2012representational,SEP-MentalRepresentation}) -- may play a key role in how we design both scientific and philosophical lines of inquiry to characterize that representation, much as we attend to how a model's target constrains the construction of and interpretation of that model in the philosophy of modeling \parencite{weisberg2013simulation,SEP-MentalRepresentation,Elliott-Graves2020What}. Because these representations are ``about" the observer's external environment (or history of its perceptions about and actions on that environment, as in memory), enabling the observer even to run predictive models based on such representations in order to plan and execute goal-directed behaviors \parencite{friston2010free}, the literature often refers to these representations as \textit{first-order representations} (FORs).

But brains do not merely represent features of the external world; they also monitor and reflect the agent's own ongoing processing, mental state, or mental structures -- including the organism's own models or representations of the world. These \textit{higher order representations} (HORs) are thus defined as being ``about" FORs \parencite{cleeremans2007conscious,Brown2019Understanding}. HORs could for example represent the signal strength in a FOR (regardless of its content) \parencite{fleming2020awareness}, or whether a FOR's content was likely externally or internally generated (i.e., real or a hallucination \parencite{lau2019consciousness, michelperceptual}), or the magnitude of noise or uncertainty present in a FOR \parencite{winter2022variance}. Note that these HORs should be distinguished on this basis from neural representations of aspects of ``higher order cognition" such as executive function or task switching, instead referring to representations that are about one's own mental state or ongoing processing. 

Such HORs receive somewhat less attention than FORs in the general literature on neural representation, their study being largely confined to those who study metacognition, meta-learning, and similar ``thinking about thinking" type approaches \parencite{Dunlosky2009Metacognition,Proust2007Metacognition}. One possible reason for this relatively smaller literature is that studying such HORs is methodologically and conceptually challenging because the processes giving rise to them may not be easily anchorable to objectively measurable observables such as behavioral reports \parencite{peters2025introspective}. This challenge has been long noted in the literature, perhaps most famously with Nisbett \& Wilson's \parencite{nisbett1977telling} observation that

\begin{displayquote}
Subjects are  sometimes (a) unaware of the existence of a stimulus that importantly influenced a response, (b) unaware of the existence of the response, and (c) unaware that the stimulus has affected the response. It is proposed that when people attempt to report on their cognitive processes, that is, on the processes mediating the effects of a stimulus on a response, they do not do so on the basis of any true introspection. Instead, their reports are based on a priori, implicit causal theories, or judgments about the extent to which a particular stimulus is a plausible cause of a given response. This suggests that though people may not be able to observe directly their cognitive processes, they  will sometimes be able to report accurately about them. (p. 231)
\end{displayquote}

\noindent This unreliability of introspective processes (which give rise to or are supported by HORs) has led some -- especially in the consciousness science community -- to hypothesize HORs to be so problematic that scientific inquiry into the topic in general may be impossible \parencite{peels2016empirical, dennett1991consciousness, schwitzgebel2008unreliability, schwitzgebel2011perplexities}. However, others -- as in the metacognition community -- have taken the stance that HORs and their associated behavioral reports may be unreliable, but that systematic patterns can nevertheless be discovered and characterized through research programs specifically designed to target their underlying processes (e.g., \cite{peters2025introspective, peters2022towards, kammerer2023what, fleming2023metacognitive, fleming2014how, Rahnev2021Visual, Peters2020Confidence}). In this piece, we build upon this second, hopeful perspective to explore how a particular kind of HOR might be constructed and scientifically studied: HORs which are about the reliability of an observer's ongoing representations and decision processes.

\subsection*{Why study higher-order representations of uncertainty?}

We focus on HORs which are specifically about noise or uncertainty in a FOR for several reasons. First, HORs about noise or uncertainty appear especially relevant for learning. For example, observers who are more ``introspectively calibrated” \parencite{fleming2014how, Maniscalco2024Optimal} –- i.e., those whose confidence better corresponds with choice accuracy and learned information –- tend to learn about their environments more quickly \parencite{meyniel2015confidence,Froemer2021Response,Hainguerlot2018Metacognitive}. This means that observers must calibrate their introspective judgments to reflect on learned environmental variables \parencite{koriat1997monitoring,Meyniel2017Brain,Meyniel2015Sense} -- even in the absence of external feedback -- to further guide the learning process itself \parencite{guggenmos2016mesolimbic,Guggenmos2022Reverse}. Studies and interventions seeking to study or even optimize learning thus may strongly benefit from a better understanding of uncertainty-related HORs.

Second, uncertainty-related HORs are also frequently invoked in inquiry into the brain's ability to distinguish reality from imagination \parencite{fleming2017self, lau2019consciousness,gershman2019generative} or generate conscious awareness \parencite{fleming2020awareness,lau2011empirical, brown2015horor, cleeremans2011radical,Michel2021Higher,Rosenthal2005Consciousness,Cleeremans2019Learning}. Specifically, Higher Order Theories (HOTs) of consciousness posit that the formation and maintenance of a HOR is both necessary and sufficient for a percept, idea, or feeling being conscious \parencite{Rosenthal2012HigherOrder}. There are several well-described HOT variants \parencite{Brown2019Understanding}. For example, Higher Order State Space (HOSS) theory \parencite{fleming2020awareness} posits that a higher order monitoring mechanism assesses the strength (and potentially reliability) of a FOR, such that if this assessment surpasses a threshold, the contents of the FOR rise into awareness. In Perceptual Reality Monitoring (PRM) theory \parencite{lau2019consciousness,michelperceptual}, it is assumed that a metacognitive mechanism estimates not only FOR strength but also whether the FOR is likely externally- or internally-sourced -- i.e., whether it is likely reflect external signals from the environment, or internally-generated imagery or noise, much like the task of a generative adversarial network (GAN) \parencite{gershman2019generative}. If the PRM mechanism fails, tagging a FOR as `real' when it was internally-generated noise, the result is theorized to be a hallucination with conscious, phenomenal quality. Higher-Order Representation of a Representation (HOROR) theory \parencite{brown2015horor} suggests that the content of a FOR is also present at the HOR level, albeit perhaps ``redescribed in a different format" \parencite{Brown2019Understanding}. A major difference among these HOT variants lies in the dimensionality and dimensions of the HOR: in HOSS, there is a signal HOR dimension (signal strength), while in PRM there are two (signal strength; reality vs. imagination) and in HOROR there are more (signal strength; reality vs. imagination; FOR content). Arbitrating these theories can benefit from more complete descriptions of HORs, including their dimensions and dimensionality. In other words, we must discover whether and how HORs may encode not only the uncertainty in an FOR, but also its strength, spatiotemporal stability, content, or any other descriptives \parencite{peters2022towards} and how read-outs of (or decision policies applied to) such HORs may drive not only metacognitive judgments but also whether the contents of an FOR are phenomenally conscious or available to behavioral report.

\section*{The components of higher-order representations of uncertainty}
Much work has sought to evaluate how HORs and the metacognitive (confidence) judgments they produce are constructed, as well as the associated neural correlates (e.g., \cite{Fleming2012Neural,shekhar2024humans,Peters2020Confidence,peters2022towards,Fleming2024Metacognition,Rahnev2021Visual,Maniscalco2016Signal}). These studies have developed a veritable zoo of potential (neural) computations giving rise to uncertainty-related HORs and confidence judgments, including charting paths forward through targeted empirical studies designed to arbitrate such theories \parencite{Rahnev2022Consensus}. 

However, less effort has been devoted to characterizing the \emph{entire} processing chain constructing HORs and confidence judgments. To be more specific, if one posits that confidence judgments result from a read-out of a HOR, then to explain those confidence judgments completely, one must describe (a) the inputs to the function \textit{generating} the HOR in the first place, (b) the function operating on those inputs, (c) the dimensions and dimensionality of the resulting HOR, and (d) the decision policy applied to the HOR to produce a confidence report. As described by Peters \parencite{peters2022towards}, nearly all perceptual metacognition and consciousness literature has confined itself to characterizing (b) from this list, with only a few studies examining deviations from the assumed standard inputs of `stimulus evidence' (e.g. \cite{winter2022variance,mamassian2022modeling,Mamassian2024CNCB}) broadly defined. A full characterization of HORs requires attention to all possible components of the metacognitive evaluation process constructing those HORs \parencite{peters2022towards}.

\begin{figure}
    \centering
    \includegraphics[width=1\linewidth]{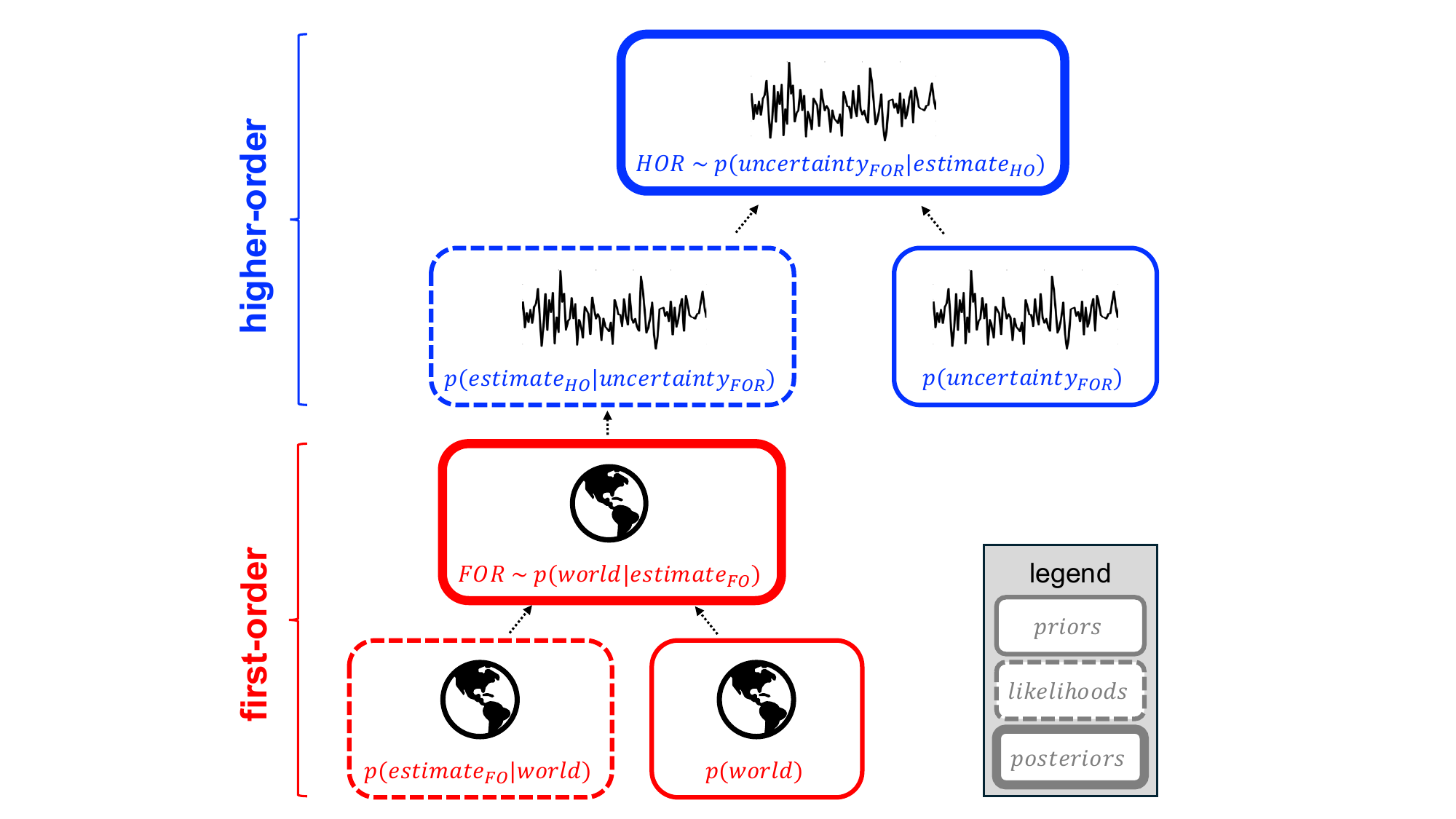}
    \caption{\textbf{Visual representation of the proposed hierarchical Bayesian process giving rise to higher-order representations (HORs) of uncertainty.} The system constructs a first-order representation (FOR; red) of the most likely state of the world  through Bayesian inference, and then again uses Bayesian inference to construct a higher-order representation (HOR; blue) about the uncertainty present in the FOR.}
    \label{fig:BayesianHORS}
\end{figure}

How can we discover the neural patterns associated with each of these components or mechanisms? A possible path forward is to specifically seek HORs of \textit{contributors} to the metacognitive estimation process -- the \textit{inputs} to the metacognitive computation as well as its outputs. That is, the metacognitive estimation process may be Bayesian-like, in which a \textit{current} estimate of uncertainty or noise is combined with the system's \textit{prior expectations} for noise under present conditions or contexts (Box~\ref{box:hor}).\footnote{In this piece we will retain the Bayesian terminology for the sake of brevity and clarity of argument. But even if one doesn't subscribe to the hypothesis that metacognition involves a Bayes-like process combining current estimates of noise with prior expectations over noise, it is reasonable to argue that a critical factor for an organism trying to evaluate its own uncertainty would before it to have some sort of `anchor' or benchmark against which to compare a current uncertainty estimate. Essentially, the system needs to be able to ask, ``Is the FOR-uncertainty I'm estimating right now large or small \textit{relative to the uncertainty I tend to experience}?" Such a comparison process necessitates the presence of some representation of FOR-uncertainty \textit{distributions}.} This proposal cleanly unifies metacognitive estimation and the construction of HORs of uncertainty with the widely successful `perception as Bayesian inference' framework \parencite{Knill1996}: the brain uses Bayesian inference to first construct a posterior estimate over the most likely state of the world given available information and prior expectations (FOR), and then again to build a posterior estimate over the most likely level of uncertainty present in that FOR (HOR) (Figure~\ref{fig:BayesianHORS}).

\refstepcounter{mybox} 

\begin{tcolorbox}[enhanced,
  colback=gray!10, colframe=black!70,
  boxrule=0.5pt, arc=4pt,
  title=Box \themybox: HOR Variants and Bayesian Structure,
  fonttitle=\bfseries,
  label={box:hor}
]
Higher-order representations (HORs) of uncertainty can be decomposed analogously to components of Bayesian inference:

\begin{center}
\begin{tabular}{@{}p{3cm}p{11cm}@{}}
\toprule
\textbf{HOR type} & \textbf{Description} \\
\midrule
\textbf{Likelihood-like} & Estimates the \emph{momentary reliability} of a current first-order representation (FOR). \\
\textbf{Prior-like} & Encodes \emph{structured expectations} about the typical noise or reliability of a target FOR along task-relevant dimension(s). \\
\textbf{Posterior-like} & Integrates both likelihood-like and prior-like components to form a \emph{composite estimate} of FOR uncertainty. \\
\bottomrule
\end{tabular}
\end{center}
By this formulation, HORs of uncertainty follow the standard Bayesian formulation: $p(uncertainty_{FOR}|estimate_{HO}) \propto p(estimate_{HO}|uncertainty_{FOR})p(uncertainty_{FOR})$.
\end{tcolorbox}

Recently, Winter \& Peters \parencite*{winter2022variance} tested the proposal that the visual system has developed prior expectations over expected uncertainty in FORs as a function of eccentricity across the visual field -- parafoveal (central) versus peripheral. They found that simple errors in the distribution of expected uncertainty could explain intriguing dissociations between \textit{actual} uncertainty in FORs (as measured by task performance accuracy) and \textit{estimated} uncertainty (as measured by subjective or metacognitive reports), and how such dissociations could be altered through task manipulations of endogenous attention. Unfortunately, though, most research, to the extent it examines (HO) representations of FOR uncertainty at all, has focused on HORs which reflect either a direct read-out of (FO) uncertainty from the perspective of the experimenter, or the \textit{result} of the estimation process as described above. In the following sections, we explore how both the \textit{instantaneous estimate} of FOR uncertainty and the \textit{distribution of expected} FOR-noise can also be considered metacognitively constructed HORs of uncertainty -- ones which have received almost no attention in the literature. 

\subsection*{The ``posterior": The experienced FOR uncertainty}

As introduced above, essentially all the work in characterizing neural representations of (HO) uncertainty has focused on discovering patterns of neural response which covary with behavioral reports about uncertainty in a task \parencite{walker2023studying}. Thus, behavioral reports provide an attractive target for revealing posterior-like HORs of uncertainty, since they reflect the result of the brain's own self-monitoring or uncertainty estimation processes. The neural correlates of these posteriors over uncertainty have been comprehensively explored by many others, so we do not extensively discuss these here; interested readers may wish to refer to \parencite{Fleming2012Neural} and \parencite{Fleming2024Metacognition} for reviews and discussion.

One can posit many possible estimation processes which may lead to such behavioral outputs \parencite{shekhar2024humans}, such as the addition of additional noise or biases at the introspective, self-monitoring level \parencite{maniscalco2012signal, maniscalco2014signal, Maniscalco2024Optimal, maniscalco2024relative, Boundy-Singer2023Confidence, mamassian2018confidence, mamassian2022modeling,Mamassian2024CNCB}. Most if not all of these implicitly assume that the metacognitive read-out reflects a direct estimate of FOR uncertainty, as with the type 2 noise posited by the signal detection theoretic measure meta-d' \parencite{maniscalco2012signal, maniscalco2014signal,Maniscalco2016signaldetectionarchitecture} and the `confidence as a noisy decision reliability estimate' (CASANDRE) model \parencite{Boundy-Singer2023Confidence}. Yet again, these models do not separate the `posterior-like' result of this estimation process from the `likelihood-like' estimate itself, as we begin to explore in this piece.

In addition to studies seeking neural correlates of the posterior-like uncertainty HORs, some model-driven neuroimaging studies have also sought to reveal neural correlates which may arbitrate between the metacognitive computations giving rise to them (e.g., \cite{peters2017perceptual}). However, while seeking neural correlates of the results of such estimation processes provides a powerful path towards understanding HORs of FOR-uncertainty, this approach does not offer a concrete focus on the \textit{components} -- or \textit{inputs} -- to such estimation processes per se \parencite{peters2022towards}. These studies are therefore limited in revealing the full heterogeneity or variety of kinds of FOR-uncertainty HORs, thus also limiting visibility into metacognitive computations themselves -- specifically, the likelihood-like and prior-like HORs of uncertainty.

\subsection*{The ``likelihood": A noisy estimate of current FOR uncertainty}

Because behaviorally-reported uncertainty is unlikely to be a direct, noiselessly-perfect readout of FOR uncertainty, instead reflecting the \textit{result} of the (potentially Bayesian) metacognitive uncertainty estimation process \parencite{winter2022variance,mamassian2024cassandre}, we would next want to characterize the likelihood function as well as this posterior. We should therefore seek to directly quantify current uncertainty in a target FOR, and then seek neural correlates which may encode this estimate: the likelihood-like HORs of uncertainty. 

Engineering approaches designed to measure uncertainty or noise in neural representations may be co-opted in service of this goal, if `pointed at' FORs. For example, the GLMsingle method \parencite{prince2022improving} couples custom hemodynamic response functions (HRFs) with regularization and cross-validation approaches to improve the reliability of beta estimates for single voxels on single trials within a task, which quantify how much a given voxel's activity is predicted by a task-relevent variable. While the goal of GLMsingle is to improve the signal-to-noise ratio of measured BOLD responses via fMRI by discarding the noise, as with any general linear model (GLM) based approach, nuisance regressors are included in the model to explicitly estimate variance \textit{not} associated with the task-relevant variables of interest -- i.e., the noise. More recently, a similar approach was developed which leverages generative models to explicitly measure noise distributions in voxel space, termed Generative Modeling of Signal and Noise (GSN) \parencite{Kay2024Disentangling}. Like GLMsingle, the goal is to improve estimates of the signal distribution in BOLD data, in this case by directly estimating the noise and then subtracting it off. 

For our purposes, one might be tempted to use GLMsingle or GSN to directly estimate voxel noise and then seek its relationship to `likelihood-like' HORs of FOR uncertainty, i.e. to discover neural signals or representations which covary with these estimated noise levels. One challenge, though, is that both GLMsingle and GSN are designed to measure voxel noise rather than FOR noise specifically, meaning that the manner in which they estimate this noise is not at all akin to how the brain might monitor its own FOR noise. In both GLMsingle and GSN, general linear models are coupled with regularization approaches which are unlikely to be directly analogous to any method employed by the brain. These same limitations are unfortunately also true for other methods specifically targeting identifying noise distributions in neural data collected via other methods, such as electrophysiology or calcium imaging \parencite{Pospisil2024Revisiting,Stringer2019HighDimensional,Williams2021Statistical}. In short, the relationship between voxelwise noise and FOR noise is as complex a the relationship between voxelwise patterns and the mental structures they represent, such that seeking neural correlates of voxelwise noise discovered through GLMsingle or GSN would not necessarily reflect HO estimates of FOR uncertainty per se.

Instead, then, one could ``read out" the uncertainty encoded in a (FO) neural representation measured via multi-unit electrophysiology using a model-based approach which explicitly dictates the relationship between neural population responses and the FORs they encode \parencite{walker2023studying}. One candidate would be probabilistic population codes, which posit that Bayesian uncertainty in a representation is specifically encoded in the gain of neural population responses \parencite{ma2006bayesian, ma2009population}. Noninvasive neuroimaging approaches have also been developed for quantifying uncertainty in FORs based on the probabilistic population coding framework: The TAFKAP method (\cite{vanbergen2021tafkap}: The Algorithm Formerly Known as PRINCE) and its predecessor PRINCE (\cite{vanBergen2015Sensory}: \textbf{Pr}obabilistic \textbf{In}ference from activity in \textbf{C}ort\textbf{e}x) both directly estimate the (FO) uncertainty in a given neural pattern along a task-relevant dimension by inverting a generative model of stimulus-evoked cortical responses. These methods have been developed to estimate probability distributions reflecting sensory uncertainty in human visual cortex during simple perceptual decision-making tasks, such as estimating the orientation/tilt of an oblique Gabor patch. The authors have reported that they can estimate this FOR uncertainty, and that observers may use knowledge of this uncertainty in their perceptual decisions: higher decoded uncertainty is related to more variable behavioral choices about the stimulus identity, i.e. lower performance and the magnitude of behavioral bias \parencite{vanBergen2015Sensory}, suggesting that human observers use knowledge of this internal uncertainty in their perceptual decision-making and can monitor fluctuations in this uncertainty from one moment (or trial) to the next. 

Importantly, though, the property measured by PRINCE and TAFKAP is FOR uncertainty from one moment to the next rather than the (likelihood-like) HOR about that uncertainty. While the authors claim that observers monitor their own FOR uncertainty and use it in behavioral decisions, their behavioral results demonstrate only that the FOR uncertainty affects decisions -- as one would expect from variations in a likelihood causing variations in a posterior judgment when a prior expectation is held constant. Because the authors did not seek to separate the observer's behavioral estimates of FOR uncertainty from actual FOR uncertainty, nor did they measure prior expectations about uncertainty, they could not assess the relationship between measured FOR uncertainty and any (likelihood- or posterior-like) HORs about it. Nevertheless, this method may show promise for future exploration of likelihood-like HOR encoding of FOR uncertainty estimates.

Despite this promising start, however, we also want to note that extending TAFKAP to areas of the brain beyond early visual cortex is also likely to be highly methodologically challenging. The response properties of early visual cortex are extremely well understood: neurons possess orientation selectivity preferences \parencite{Kamitani2005Decoding,Haynes2005Predicting,Brouwer2011CrossOrientation,Serences2009Estimating,kay2008identifying,Jehee2012Perceptual}, and individual neurons' activity exhibit well characterized noise correlations across trials \parencite{Smith2008Spatial,Goris2014Partitioning}. This deep knowledge of visual cortex response properties makes it possible to develop the generative models on which TAFKAP's success relies. Unfortunately, though, response properties of other FORs are less well characterized -- for example, selectivity is more mixed in later visual processing areas such as inferiortemporal cortex (e.g., \cite{Chang2021Explaining,Bao2020Map}). Discovering the coding properties of ``higher" level FORs beyond early visual cortex is a massive undertaking in its own right. As such response properties are revealed, however, it may be possible to marry TAFKAP-like methods with behavioral metrics of HOR-derived uncertainty estimates.

\subsection*{The ``prior": Expectations about FOR uncertainty}

As with any Bayesian model, there is one component remaining to discuss: the prior, or the expectations about FOR uncertainty that the observer has built through experience \parencite{Series2013Learning}. How might we go about understanding such \textit{expected FOR-noise distribution} neural HORs? 

To characterize any novel distribution, one might start by simply taking samples. But we cannot take samples from the posterior (confidence reports or associated neural correlates). To begin measuring \textit{expected noise-distribution} HORs, one might instead employ psychophysical measures such as those previously used to recover priors about environmental variables used in FORs -- for example in perception \parencite{Stocker2006Noise,Girshick2011Cardinal,Peters2015Smaller,Adams2004Experience,Series2013Learning,Odegaard2016Brains,Odegaard2015Biases}. In these studies, behavioral measurements are first used to measure the percept (the Bayesian posterior) of e.g. orientation, speed, or object heaviness; then, through manipulating the environmental noise present in the stimuli, one can decompose the combined estimate (posterior distribution) into a Bayesian combination of the instantaneous, noisy estimate of the environmental variable of interest (likelihood) and the prior used by the observer. Such studies have revealed priors across environmental variables such as spatial location \parencite{Odegaard2015Biases}, motion speed \parencite{Stocker2006Noise}, visual contour orientation \parencite{Girshick2011Cardinal}, light source location \parencite{Adams2004Experience}, and even tendency to bind multisensory stimuli \parencite{Odegaard2016Brains}, for example. Used in conjunction with metacognitive judgments about FOR-uncertainty or confidence, this approach may provide insight into task-specific or context-conditioned distributions of expected noise, which could then be used to drive discovery of their neural correlates. However, contextually-conditioned distributions of expected noise then would be confined to a particular variable or task of interest, which -- while interesting and fruitful in the context of certain observable variables in the environment such as contour orientation, object density, and so on -- will not give us an understanding of the \textit{full} landscape of FOR-noise distributions in the brain, or how they are learned by the system. 

Instead of using behavior alone, then, another possibility is to directly sample from the neural prior-like uncertainty HOR itself as it varies across tasks, context, or time using neuroimaging approaches. Note that this approach requires specifically that we sample from the prior-like HOR, not just sample the distribution of noise in the brain across task, context, or time; a resting state scan, for example, would be insufficient. Instead, sampling from the prior-like HOR directly would require identifying a target, task-relevant dimension of the FOR about which uncertainty may be estimated, and being able to track how the brain builds HORs about this dimension so as to eventually map it back to brain response.

\begin{figure}[ht]
    \centering
    \fbox{\includegraphics[width=.95\textwidth]{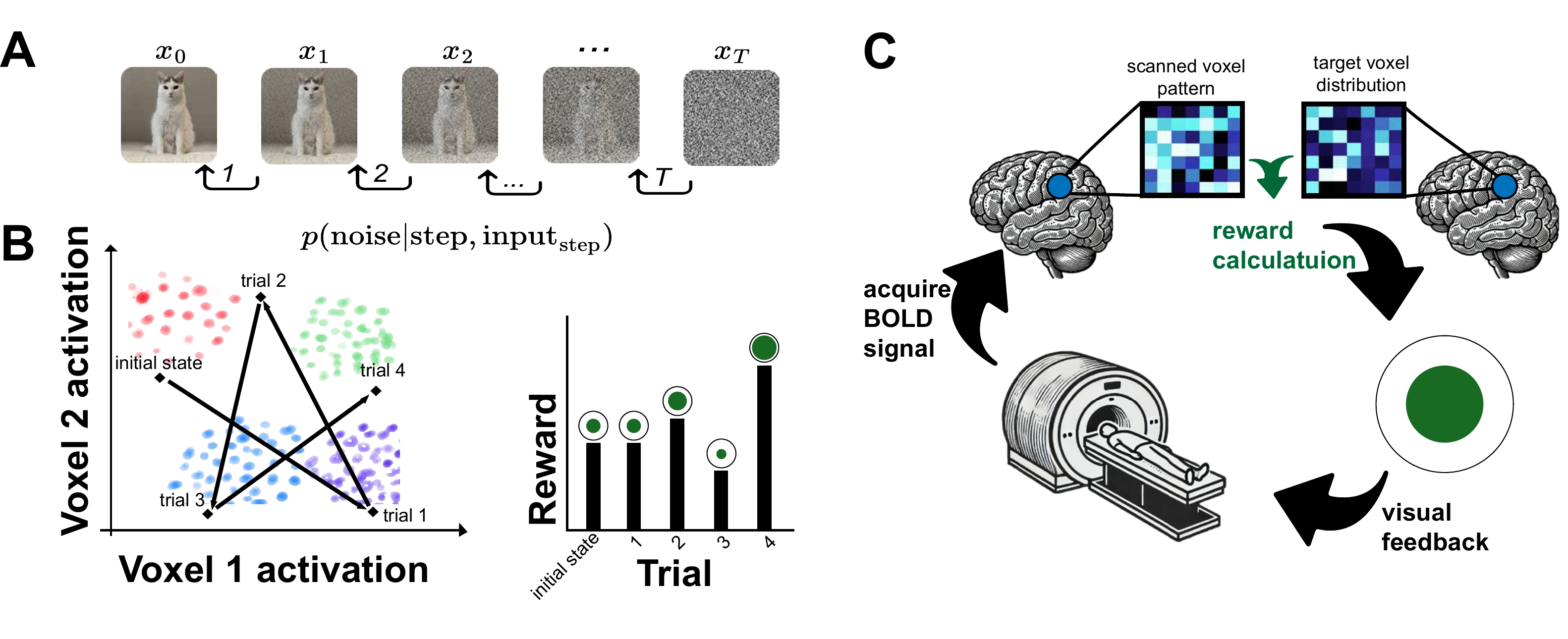}}
    \caption{\textbf{Cartoons showing the denoising process learned by diffusion models and the closed-loop real-time neurofeedback training procedure.} (A) Denoising diffusion models are trained to learn distributions of pixel noise, conditioned on the denoising step and input image $x_T$, i.e. $p(noise|step, input_{step})$, in order to denoise the input image such that a new image $x_0$ from the target distribution can be produced. (B) We have recently hypothesized \parencite{asrariPetersTechnical} that this denoising process is undertaken by brains in order to navigate through possible neural patterns in search of a refined goal state, which is likely accomplished through reinforcement learning (RL) in environments where the goal state is not known to the observer. Denoising -- or uncertainty reduction -- provides a natural candidate mechanism that the brain is equipped to attempt even when the specific goal state is totally unknown. In decoded neurofeedback (DecNef), the observer seeks states which minimize the difference between the current state and the \textit{distribution} of target states, and the degree of match is displayed to the observer as a visualization of computed reward. (C) The closed-loop DecNef procedure involves human subjects learning to denoise their own brain states through RL. Neural response patterns (blood oxygen level dependent [BOLD] signal) are acquired in a given region of interest using functional magnetic resonance imaging (fMRI), compared to a target neural pattern (defined by previous activity patterns), and the degree of similarity between current and target neural state is displayed back to the human participant in the form of a visual feedback circle.}
    \label{fig:introDenoisingCartoon}
\end{figure}

We recently proposed an approach to achieving this prior-like HOR sampling: the Noise Estimation through Reinforcement-based Diffusion (NERD) model, which uses generative artificial intelligence (genAI) algorithms designed specifically to learn noise distributions (in service of iteratively `denoising' patterns -- e.g., images, music, or text -- to produce a target pattern) from empirical neuroimaging data collected from humans who learned to do a similar task (to iteratively `denoise' their neural activity patterns to achieve a target pattern) \parencite{azimi2024diffusion,asrariPetersTechnical}. After training, NERD possesses in its model architecture and fitted parameters a representation of the \textit{distribution} of voxel pattern noise learned across the task (i.e., $p(noise|step, input_{step})$; Fig. \ref{fig:introDenoisingCartoon}a), such that it in essence represents the results of `sampling the noise'. 

To map the learned noise-distribution HORs back to neural activity patterns, we sought a set of sampled points in neural state space which could be directly compared to sampled points in the NERD model. This required that the task learned by NERD, the way NERD learns the task, and the data used to train NERD to do this task, need be as analogous as possible to the task and data structure for the human data. In other words, the model must have learned about its own prior-like noise distributions in a similar way as a target human did, and using similar data. 

To accomplish this goal, we used data from a Decoded Neurofeedback (DecNef) (Fig. \ref{fig:introDenoisingCartoon}b,c) task, in which human subjects learned to alter the patterns of their own brain response in order to achieve a target goal pattern. Functional magnetic resonance imaging (fMRI) DecNef combines real-time fMRI with multivariate pattern analysis to allow individuals to regulate complex brain activity patterns voluntarily \parencite{laconte2011decoding, Cortese2021}. Machine learning algorithms are trained to decode specific (FO) mental states from participants' fMRI activation patterns; by providing continuous feedback on their ongoing mental state, DecNef trains participants to modulate the associated brain activity patterns in target regions of interest (ROIs) \parencite{Shibata2011, watanabe2017advances}, specifically along a task-relevant dimension.  

We proposed that one way human subjects can solve the DecNef task is by learning about the uncertainty in their own neural representations, and then navigating this distribution of noise -- essentially `denoising' neural patterns -- to achieve their target goal \parencite{azimi2024diffusion,asrariPetersTechnical}. We posited this in part because in DecNef the goal state is entirely unknown to the subject, and the subject receives no explicit instructions on what the target pattern should be. It has previously been proposed that human subjects learn to achieve target patterns through reinforcement learning (RL), because the DecNef procedure involves a pretrained classifier comparing the current brain state to the target brain state and then displaying the discrepancy to the user in the form of visual feedback reward \parencite{Shibata2011}. Because no instructions are given to the subject other than `maximize the reward you can achieve', we hypothesized that the brain may `solve' DecNef through engaging a procedure it \textit{does} know how to do: uncertainty reduction. Uncertainty reduction is a core capacity for all biological brains which may guide perception, action, curiosity, and information seeking \parencite{gottlieb2018towards, gottlieb2013information,DeRidder2014Bayesian,Friston2017Active}. 

We thus suggested that NERD -- a denoising diffusion model trained with RL algorithms -- could provide a powerful framework for capturing the brain's process of learning about its own noise in service of minimizing uncertainty. We then trained NERD on an existing DecNef dataset, and projected it into HO space (rather than voxel noise space) using dimensionality reduction techniques that are often used to link neural patterns with representations (mental structures) \parencite{Bishop2006, Schneider2023Learnable, Steinmetz2021Neuropixels, Stringer2019HighDimensional,Pearson1901,VazquezGarcia2024Review}. With this approach, we found that the lower-dimensional prior-like uncertainty HORs discovered by NERD could indeed capture individual variation in humans' ability to solve the DecNef task. We believe that the noise distributions learned by NERD under these conditions could thus provide a window into these uncertainty priors as component inputs to the construction of posterior-like HORs about uncertainty. These findings suggest that the brain not only tracks immediate uncertainty, but also builds expectations about uncertainty patterns which likely guide how we learn and make decisions, supporting the proposal developed here that HORs about uncertainty are constructed through a Bayesian-like process.

\section*{Summary and final thoughts}

Here we've discussed varieties of higher-order representations (HORs), with specific attention to the component ingredients that serve as inputs to the metacognitive processes constructing HORs. We proposed that HORs about uncertainty may be constructed according to a Bayesian-like process, such that reported metacognitive estimates of uncertainty are posterior-like, reflecting a combination of prior-like and likelihood-like HO estimates of uncertainty. We then discussed how various approaches, from the GLMsingle \parencite{prince2022improving} and GSN \parencite{Kay2024Disentangling} approaches to TAFKAP \parencite{vanBergen2015Sensory,vanbergen2021tafkap} and our newly-developed NERD model \parencite{azimi2024diffusion,asrariPetersTechnical} may provide exciting paths forward towards studying these component ingredients of uncertainty HORs. 

We believe that formulating the construction of uncertainty HORs as Bayesian-like can add crucial clarity to our understanding of the variety of uncertainty-based HORs computed by the brain, as well as their neural correlates and generating computations. We also hope this discussion may also inspire and enable a systematic exploration of HORs in general, beyond those which are about FOR uncertainty, providing crucial insight into the complex relationships between neural patterns and the kinds of mental states they represent.

\section*{Conflict of Interest Statement}
M.A.K.P. is a consultant for the for-profit entity Conscium, Inc., which seeks to pioneer safe, efficient artificial intelligence and which played no role in this project's conceptualization, analyses, interpretation, or writing. The authors declare no conflicts of interest.

\section*{Acknowledgments}
This project was supported in part by a fellowship (to M.A.K.P.) from the Canadian Institute for Advanced Research Program in Brain, Mind, \& Consciousness and a grant from the Templeton World Charity Foundation (``An adversarial collaboration to empirically evaluate higher-order theories of consciousness", to M.A.K.P.). The funding sources had no role in the design, implementation, or interpretation of the work presented here.

\newpage
\printbibliography
\end{document}